\documentclass[aps,twocolumn,prl,showpacs,superscriptaddress]{revtex4-2}

\usepackage{bm}
\usepackage{graphicx}
\usepackage{amsmath}
\usepackage{amssymb}
\usepackage{color}
\usepackage{wasysym}
\usepackage{enumerate}
\usepackage[colorlinks=true,%
            citecolor=blue,%
            linkcolor=red,%
            urlcolor=blue]{hyperref}
\usepackage{ifthen}
\usepackage{xspace}
\usepackage[caption=false]{subfig}
\def\today{August 9, 2023; Revised: January 9, 2024} 
\begin{document}

\title{Shortcuts of freely relaxing systems using equilibrium physical 
observables}

\author{Isidoro Gonz\'alez-Adalid Pemart\'{\i}n}\affiliation{Departamento 
de F\'{\i}sica Te\'orica, Universidad Complutense, 28040 Madrid, Spain}%

\author{Emanuel Momp\'o}\affiliation{Departamento de Matem\'atica Aplicada, 
Grupo de Din\'amica No Lineal, Universidad Pontificia Comillas, Alberto 
Aguilera 25, 28015 Madrid, Spain}%
\affiliation{Instituto de Investigaci\'on Tecnol\'ogica (IIT), Universidad 
Pontificia Comillas, 28015 Madrid, Spain}%

\author{Antonio Lasanta}\email[Corresponding author: ]{alasanta@ugr.es}%
\affiliation{Departamento de \'Algebra, Facultad de Educaci\'on, Econom\'ia y 
Tecnolog\'ia de Ceuta, Universidad de Granada, Cortadura del Valle, s/n, 
51001 Ceuta, Spain}%
\affiliation{Nanoparticles Trapping Laboratory, Universidad de Granada, Granada, Spain}%

\author{V\'{\i}ctor Mart\'{\i}n-Mayor}\affiliation{Departamento de F\'{\i}sica 
Te\'orica, Universidad Complutense, 28040 Madrid, Spain}%
\affiliation{Instituto de Biocomputaci\'on y F\'{\i}sica de Sistemas 
Complejos (BIFI), 50018 Zaragoza, Spain}%

\author{Jes\'us Salas}
\affiliation{Departamento de Matem\'aticas, Universidad Carlos III de Madrid, 
28911 Legan\'es, Spain}%
\affiliation{Grupo de Teor\'{\i}as de Campos y F\'{\i}sica Estad\'{\i}stica, 
Instituto Gregorio Mill\'an, Universidad Carlos III de Madrid, Unidad 
Asociada al Instituto de Estructura de la Materia, CSIC, Spain}%

\date{\today}

\begin{abstract}
Many systems, when initially placed far from equilibrium, exhibit surprising
behavior in their attempt to equilibrate. Striking examples are the Mpemba
effect and the cooling-heating asymmetry. These anomalous behaviors can be
exploited to shorten the time needed to cool down (or heat up) a
system. Though, a strategy to design these effects in mesoscopic systems is
missing. We bring forward a description that allows us to formulate such
strategies, and, along the way, makes natural these paradoxical
behaviors. 
In particular, we study the evolution of macroscopic physical observables of 
systems freely relaxing under the influence of one or two instantaneous 
thermal quenches.
The two crucial ingredients in our approach are timescale separation and a 
nonmonotonic temperature evolution of an important state function. We argue 
that both are generic features near a first-order transition.
Our theory is exemplified with the one-dimensional Ising model in a magnetic 
field using analytic results and numerical experiments.
\end{abstract}

\maketitle

%%%%%%%%%%%%%%%%%%%%%%%%%%%%%%%%%%%%%%%%%%
%
%    SECTION: Introduction
%
%%%%%%%%%%%%%%%%%%%%%%%%%%%%%%%%%%%%%%%%%%
%\noindent
\paragraph{Introduction.}
Controlling and predicting relaxation processes far from equilibrium is 
still an open task. 
In spite of historical advances mostly achieved along the 20th 
century~\cite{Nyquist1928,Zwanzig1965,onsager1931,onsager19312,kubo1966,%
marconi2008}, with the exception of some cases in molecular gases 
\cite{evans20081,evans20082}, we lack a general theory beyond the linear 
response theory and fluctuation theorems allowing us to manage transient 
regimes and, in particular, optimize relaxation times of a freely evolving 
system between two desired states~\cite{Seifert2012}. Recent progress 
unraveling anomalous relaxation processes in out-of-equilibrium systems 
points in that direction.

An  outstanding example is the Mpemba effect 
(ME)~\cite{jeng2006,mpemba1969,Bechhoefer2021}. Put instantly two 
systems---identical but for their different initial temperatures---in contact 
with a thermal bath at a colder-than-both temperature. The ME happens 
when the initially hotter system cools faster than the system that was 
initially closer to equilibrium.

In Markovian systems, the ME can be well understood using a spectral 
decomposition and diminishing or canceling slow-decaying modes for the sake 
of enhancing the fast ones. This has been done both in 
classical~\cite{LuRaz2017,klich2019,kumar2020,walker2021,biswas2023,%
Schwar2022,kumar22,Vadakkayil2021} and open quantum 
systems~\cite{carollo2021,nava2019,Chatterjee20232}. Meanwhile, in systems 
where spectral methods are not applicable, other strategies can be 
used for controlling fast and slow evolution using macroscopic observables. 
Namely, energy nonequipartition in water~\cite{gijon2019}, a particular 
condition in kurtosis in granular 
gases~\cite{lasanta2017,torrente2019,mompo2021}, and correlation length 
in spin glasses~\cite{BaityJesi2019}. Furthermore, other strategies 
using several quenches have been shown to be useful in attaining a speedup 
in relaxation times: preheating protocols~\cite{gal2020}, taking advantage 
of magnetic domains growth when a large number of degrees of freedom 
near phase transitions are present in the system and timescale separation is 
not possible~\cite{pemartin2021}, or different control 
techniques~\cite{guery2022,chittari2023}. 

We shall not overlook that achieving a speedup is not a trivial task. As 
it has been experimentally shown, the Kovacs effect prevents fast 
relaxation when using two quenches in a naive way, either for heating or 
cooling~\cite{Kovacs1964,Kovacs1979,Prados2014,militaru2021}. What is 
even more surprising is that, recently, another anomaly which was verified 
both theoretically and experimentally has been found: far from equilibrium, 
there can appear an asymmetry between equidistant and symmetric heating and 
cooling processes~\cite{lapolla2020,ibanez2023}. Even more fundamental is 
that, using reciprocal relaxation processes between two fixed temperatures, 
the asymmetry is also found \cite{ibanez2023}. This has been successfully 
explained using the so-called ``thermal kinematics''~\cite{ibanez2023} 
based on information geometry~\cite{crooks2007, Ito2020Stochastic}. 

Here, we aim to control the out-of-equilibrium evolution of a system
\emph{solely} relying on its in-equilibrium physical observables and
on the spectral decomposition of the dynamic-generating matrix.
Our focus is on observable dynamics: clever computational speedups lacking 
an experimental counterpart (think, e.g., of Ref.~\cite{wolff:89}) are excluded.
The physical interpretation is straightforward. By identifying the
slowest-decaying physical observables, we are able to project the
system under study onto the faster ones in order to speed up the total
relaxation of the system. Near first order phase transitions, the desired 
fast relaxation can be achieved
choosing the appropriate initial condition, or, previous to the final
relaxation, by briefly heating or cooling the system. To showcase
this, we use the antiferromagnetic (AF) 1D Ising model with a
magnetic field. 

%%%%%%%%%%%%%%%%%%%%%%%%%%%%%%%%%%%%%%%%%%
%
%    SECTION: Theory / proposed approach
%
%%%%%%%%%%%%%%%%%%%%%%%%%%%%%%%%%%%%%%%%%%
\paragraph{Theoretical framework.}
We shall model the microscopic dynamics in contact with a thermal bath at
temperature $T_b$ through a Markov dynamics with continuous
time~\cite{levin2017markov}, obtained as the continuous limit of some
discrete-time Markov chain (in our case, heat-bath
dynamics~\cite{sokal1997functional}). 

There are two complementary
viewpoints on Markov dynamics. Either one considers the time evolution
of the probability distribution function (the so-called strong form of
the associated stochastic differential equation), or one focuses on
the time evolution of observable magnitudes (the weak
form)~\footnote{These two viewpoints remind the Heisenberg
  and \protect{Schr\"odinger} pictures of time evolution in quantum
  mechanics.}.  While the strong form has been emphasized in recent
work~\cite{chittari2023}, we shall privilege the very insightful
weak-form approach. We  briefly recall now the main ingredients of both 
approaches (see~\cite{levin2017markov,supplemental} for details).

Let \(\Omega\) be the set of
all possible states of a system~\footnote{For a chain of $N$ Ising spins, the
number of states is $|\Omega|=2^N$. For $|\Omega|=\infty$, the natural
mathematical framework is functional analysis (rather than linear
algebra).}. The strong form of the dynamics focus on the master equation 
for $P_{\boldsymbol{y}}^{(t)}$, the probability of finding the system in 
the microscopic state $\boldsymbol{y}$ at time $t$:
\begin{equation}\label{eq:master-equation}
\frac{\mathrm{d}P_{\boldsymbol{y}}^{(t)}}{\mathrm{d} t} \;=\; 
\frac{1}{\tau_0}\,\sum_{\boldsymbol{x}\in\Omega} 
P_{\boldsymbol{x}}^{(t)} R_{\boldsymbol{x},\boldsymbol{y}}\,,
\end{equation}
where $R_{\boldsymbol{x},\boldsymbol{y}}/\tau_0$ is the
probability per unit time for the system to jump from state
\(\boldsymbol{x}\) to state \(\boldsymbol{y}\) when subject to a
thermal bath with temperature \(T_b\) ($\tau_0$ is a fixed
time unit). Setting the diagonal term as
$R_{\boldsymbol{x},\boldsymbol{x}}=
-\sum_{\boldsymbol{y}\in\Omega \setminus \{\bm{x}\}}
R_{\boldsymbol{x},\boldsymbol{y}}$ ensures the conservation of the
total probability. The master equation can be solved  by expressing
the initial probability $\bm{P}^{(t=0)}$ as a linear combination of the 
left eigenvectors of the matrix $R$ (see, 
e.g.,~\cite{levin2017markov,supplemental}), but this would take us too far 
off field. Instead, we wish to focus on the weak form
of the dynamics, for which there are two crucial mathematical ingredients. 

Our first ingredient is the inner product between two
observables, $\mathcal{A}$ and $\mathcal{B}$ (i.e. two mappings from
\(\Omega\) to the real numbers). 
Let $\mathbb{E}^T[\mathcal{A}]=\sum_{\boldsymbol{x}\in\Omega}\, 
\pi_{\boldsymbol{x}}^T\,\mathcal{A}(\boldsymbol{x})$ be the equilibrium 
expected value of $\mathcal{A}$ at temperature $T$ ($\pi_{\bm{x}}^T$ is the
Boltzmann weight for state $\bm{x}$). The
inner product of $\mathcal{A}$ and $\mathcal{B}$ is defined at the 
bath temperature:
\begin{equation}\label{eq:producto-escalar}
  \langle \,\mathcal{A}\, |\, \mathcal{B} \,\rangle \;:=\; 
\mathbb{E}^{T_b}[\mathcal{A}\, \mathcal{B}] \;=\;
\sum_{\boldsymbol{x}\in\Omega}\, \pi_{\boldsymbol{x}}^{T_b}\, 
\mathcal{A}(\boldsymbol{x})\, \mathcal{B}(\boldsymbol{x})\,.
\end{equation}
In particular, let \(\boldsymbol{1}\) be the constant observable such
that \(\boldsymbol{1}(\boldsymbol{x})=1\) for any state
$\boldsymbol{x}$. Hence, for any observable  
${\mathcal A}$, \(\,\langle \boldsymbol{1}\,|\,\mathcal{A}\,\rangle =
\mathbb{E}^{T_b}[\mathcal{A}]\), while the equilibrium
variance at temperature \(T_{b}\) is 
$\langle \mathcal{A}^\bot\,|\,\mathcal{A}^\bot\,\rangle$, where 
$\mathcal{A}^\bot:=\mathcal{A}-\boldsymbol{1} 
\mathbb{E}^{T_b}[\mathcal{A}]$ accounts for the \emph{fluctuations} 
of $\mathcal{A}$ from its expected value at \(T_b\). Furthermore, 
the fluctuation-dissipation theorem tells us that 
\begin{equation}\label{eq:FDT}
\left.T^2 \, 
\frac{\mathrm{d} \mathbb{E}^{T}[\mathcal{A}]}{\mathrm{d} T}
\right|_{T=T_b} \;=\; \langle\, \mathcal{A}^\bot\, |\, 
\mathcal{E}\,\rangle
\end{equation}
($\mathcal{E}$ is the energy, and 
$\langle \mathcal{A}^\bot\, |\, \mathcal{E}\rangle\, =
\mathbb{E}^T[\mathcal{A}\,\mathcal{E}]\,-\, 
\mathbb{E}^T[\mathcal{A}]\,\mathbb{E}^T[\mathcal{E}]$).

Our second crucial ingredient is the operator ${\cal R}$, 
that generates the time evolution of observables:
${\cal R}[\mathcal{A}] (\boldsymbol{x})=
\sum_{\boldsymbol{y}\in\Omega} R_{\boldsymbol{x},\boldsymbol{y}}
\mathcal{A}(\boldsymbol{y})$ [the matrix $R$ was defined in 
Eq.~\eqref{eq:master-equation}].  In particular, 
${\cal R} [\boldsymbol{1}](\boldsymbol{x})=0$ for all $\boldsymbol{x}$
due to probability conservation (hence, $\boldsymbol{1}$ is an eigenfunction:
${\cal R} [\boldsymbol{1}]=0 \cdot \boldsymbol{1}$).

Detailed balance implies that ${\cal R}$
is self-adjoint with respect to the inner product~\eqref{eq:producto-escalar}.
For any  $\mathcal{A}$ and $\mathcal{B}$
\begin{equation}\label{eq:detailed-balance}
  \langle \, {\cal R}[\mathcal{A}] \, |\,  \mathcal{B} \, \rangle=
  \langle \, \mathcal{A}\, |\, {\cal R}[\mathcal{B}]  \, \rangle\,.
\end{equation}
It follows that we can find an orthonormal basis of the space of observables 
with finite variance
\((\boldsymbol{1},\mathcal{O}_2^{b},\mathcal{O}_3^{b},\ldots)\),
in which the $\mathcal{O}_k^{b}$ are all eigenfunctions
\({\cal R}[\mathcal{O}_k^{b}] =\lambda_k
\mathcal{O}_k^{b}\)~\footnote{Superscripts $b$  are twofold. 
  They stress the fact that the observables \(\mathcal{O}^{b}_k\) 
  are tied to the bath temperature \(T_{b}\), and that detailed balance,
  Eq.~\eqref{eq:detailed-balance}, is only fulfilled for the scalar product 
  at temperature \protect{$T_{b}$}.
}.  We order the basis in such a way that 
$0=\lambda_1>\lambda_2\geq\lambda_3\geq\ldots$. Take now an arbitrary
starting distribution function $\bm{P}^{(t=0)}$ at time $t=0$. The
expected value of \emph{any} finite-variance observable $\mathcal{A}$ at time
$t>0$ is
\begin{eqnarray}\label{eq:spectral_decomposition}
  \mathbb{E}_{t}[\mathcal{A}] &=& \mathbb{E}^{T_b}[\mathcal{A}] + 
\sum_{k\geq2} \alpha_k^{(t=0)} \beta_{k}^{\mathcal{A}}\, 
              \mathrm{e}^{-|\lambda_k|t/\tau_0}\,,\\
\beta_{k}^{\mathcal{A}} &=& \langle \, {\mathcal O}_k^b\,|\,
              \mathcal{A}\,\rangle\,, \quad 
\alpha_k^{(t=0)}\;=\;\sum_{\boldsymbol{x}\in\Omega}\,
 P_{\boldsymbol{x}}^{(t=0)}\,{\cal O}_k^{b}(\boldsymbol{x})\,.   \quad  
\end{eqnarray}
As long as the system shows separation of timescales
(i.e. \(|\lambda_2| < |\lambda_3|\)),
Eq.~\eqref{eq:spectral_decomposition} gives rise to a hierarchy of
physical magnitudes, with \(\mathcal{O}_2^{b}\) having the slowest
decay. If we are able to find an initial setup such that
$\alpha_2^{(t=0)}=0$---all $\alpha_k^{(t=0)}$ are independent of the
observable $\mathcal{A}$ under consideration---then, provided that
$\beta_2^\mathcal{A}\neq 0$, its expected
value will benefit from an exponential speedup in the evolution
toward its equilibrium value
\(\mathbb{E}^{T_{b}}[\mathcal{A}]\).

It is important to notice that the initial setup that will be used to
speed up the system---coded in the starting probability
$\bm{P}^{(t=0)}$---is not restricted to an equilibrium
state. However, our interest lies on equilibrium states because they are
easier to control (e.g., experimentally). For instance,
using an equilibrium initial condition, the requirement for a speedup
would be met if we can find a temperature \(T\neq T_{b}\) such that
\(\alpha_2^{(t=0)}\equiv
\mathbb{E}^{T}[\mathcal{O}_2^{b}]=0\). We show below how to find 
(or force) this condition, explaining along the way
different anomalous behaviors.

%%%%%%%%%%%%%%%%%%%%%%%%%%%%%%%%%%%%%%%%%%
%
%    SECTION: Experimental benefits
%
%%%%%%%%%%%%%%%%%%%%%%%%%%%%%%%%%%%%%%%%%%
\paragraph{Experimental suitability.} 
If the number of states is large enough, the spectral 
decomposition~\eqref{eq:spectral_decomposition} might not be practical.
In particular, the slowest decaying observable \(\mathcal{O}_2^{b}\)
might be unknown (or, even if known, its experimental measurement may
present difficulties). Nonetheless,  in some situations there may
be a simple way out.

Specifically, let us consider the neighborhood of a first-order phase 
transition at zero temperature separating two ground states with different 
symmetries (we consider a specific example below). Let 
$\mathcal{M}_{w}$ be the
order parameter of the unstable phase (the suffix $w$ stands 
for \emph{wrong}). If symmetries are such that
$\mathbb{E}^T[\mathcal{M}_{w}]=0$ for all $T$, then there 
are good chances that
\((\mathcal{M}_{w}^2)^\bot\) will be a nice proxy for 
\(\mathcal{O}_2^{b}\). Indeed, at phase coexistence, slow observable 
dynamics (OD) often stems from metastability~\cite{parisi:88}. We consider 
\((\mathcal{M}_{w}^2)^\bot\),  the fluctuating part of 
\(\mathcal{M}_{w}^2\), to mimic the behavior of 
\(\mathcal{O}_2^{b}\): 
$0=\langle \boldsymbol{1}\,|\, \mathcal{O}_2^{b}\rangle=
\mathbb{E}^{T_{b}} [\mathcal{O}_2^{b}]$.  

We expect
$1\approx \langle (\mathcal{M}_{w}^2)^\bot|\, 
\mathcal{O}_2^{b}\rangle\,/\,\langle 
(\mathcal{M}_{w}^2)^\bot|\, (\mathcal{M}_{w}^2)^\bot 
\rangle^{1/2}$ for any good proxy: in geometrical terms, the angle between
\(\mathcal{O}_2^{b}\) and \((\mathcal{M}_{w}^2)^\bot\) defined 
by the scalar product~\eqref{eq:producto-escalar} will be small.

\begin{figure}[tb]
\includegraphics[width=8.6cm]{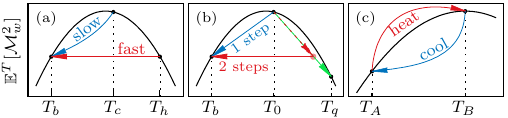}
\caption{Staging the anomalous effects. (a) The Mpemba effect. (b) Preheating 
for faster cooling. (c) Asymmetry of heating and cooling processes.}
\label{fig:anomalous_effects}
\end{figure}

\begin{figure}[htb]
\includegraphics[width=8.6cm]{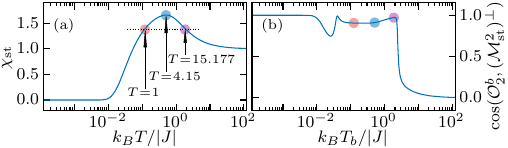}
\caption{Validation of $(M^2_\text{st})^\bot$ as a proxy for the slowest 
decaying observable \(\mathcal{O}_2^{b}\).
(a) \(\chi_{\text{st}}\) as computed for $N\to\infty$ vs. $k_BT/|J|$, 
see Eq.~\eqref{eq:def-chi_st-C1},  has a maximum at 
\(T_{b}^* \approx 4.15\). The curves for $N=8$ and $N\to\infty$ are 
hardly distinguishable at this scale. 
(b) For $N=8$, the cosine of the angle between $(\mathcal{M}_\text{st}^2)^\bot$
and \(\mathcal{O}_2^{b}\) is close to 1 for a wide range of bath 
temperatures. The colored dots in (a) and (b) indicate
the three temperatures that we mostly use to demonstrate unconventional OD.}
\label{fig:ansatz_check}
\end{figure}

The crucial point leading to unconventional OD effects is
that $\mathbb{E}^T[\mathcal{M}_{w}^2]$ is a 
\textit{nonmonotonic function} of $T$, and we can find a bath temperature  
\(T_{b}^*\) such that
$\text{d}\, 
\mathbb{E}^{T}[\mathcal{M}^2_{w}]/\text{d}T|_{T=T_{b}^*} = 0$
(see Figs.~\ref{fig:anomalous_effects} and~\ref{fig:ansatz_check}). 
This maximum generates surprising OD.

The behavior in Figs.~\ref{fig:anomalous_effects} and~\ref{fig:ansatz_check}
is generic because $\mathbb{E}^T[\mathcal{M}_{w}^2]$
is proportional to the susceptibility of $\mathcal{M}_{w}$ with 
respect to its conjugate field. Now, this susceptibility vanishes when
$T\to 0$ (because $\mathcal{M}_{w}$ is the order parameter
for the unstable phase) while, for large enough $T$,
all susceptibilities decrease as $T$ grows.

%%%%%%%%%%%%%%%%%%%%%%%%%%%%%%%%%%%%%%%%%%
%
%    SECTION: Explanation of anomalous effects
%
%%%%%%%%%%%%%%%%%%%%%%%%%%%%%%%%%%%%%%%%%%
\paragraph{The basic observation.} Consider the spectral 
decomposition~\eqref{eq:spectral_decomposition} when the starting
distribution $\bm{P}^{(t=0)}$ is the Boltzmann weight for some temperature 
$T^*\neq T_{b}$. We can approximate coefficient $\alpha^{(t=0)}_{2}$ as
\begin{equation}\label{eq:astucia}
\alpha^{(t=0)}_{2}\approx \frac{1}{\Lambda}
\Big(\mathbb{E}^{T^*}[\mathcal{M}^2_{w}]\ -\ 
\mathbb{E}^{T_{b}}[\mathcal{M}^2_{w}]\Big)\,,
\end{equation}
with $\Lambda=\langle (\mathcal{M}_{w}^2)^\bot|\, 
(\mathcal{M}_{w}^2)^\bot \rangle^{1/2}$ a relatively uninteresting 
constant fixed by the bath temperature~\footnote{Recall that 
   \((\mathcal{M}_{w}^2)^\bot = 
   \mathcal{M}_{w}^2 - 
   \boldsymbol{1}\mathbb{E}^{T_{b}}[\mathcal{M}_{w}^2]\) mimics 
   \(\mathcal{O}_2^{b}\) in the sense that 
   \(\mathbb{E}^{T_{b}}[\mathcal{O}_2^{b}]=0\). 
   After accounting for normalization, 
   \(\alpha^{(t=0)}_{2}\approx\Lambda^{-1}\,
   \mathbb{E}^{T_{b}}[(\mathcal{M}_{w}^2)^\bot]\), which leads 
   to Eq.~\eqref{eq:astucia}.
}. In the more general case, $\mathbb{E}^{T^*}[\mathcal{M}^2_{w}]$ 
should be traded with $\mathbb{E}_{t=0}[\mathcal{M}^2_{w}]$ in 
Eq.~\eqref{eq:astucia}. Several anomalous effects can be better understood 
from this simple observation.

\paragraph{The Markovian Mpemba effect~\cite{LuRaz2017}.} Consider a bath 
at temperature $T_{b}$ and two other temperatures, $T_{c}$ 
and $T_{h}$ such that \(T_{b}<T_{c}<T_{h}\), 
chosen to have expected values of
$\mathcal{M}^2_{w}$ as shown in Fig.~\ref{fig:anomalous_effects}(a). 
In the view of Eq.~\eqref{eq:astucia},
it is clear that $\alpha_2^{h}=0$, while $|\alpha_2^{c}|>0$.
This means that, provided that we start from a system in equilibrium at 
$T_{h}$,
any observable $\mathcal{A}$ with $\beta_2^{\mathcal{A}}\neq 0$ 
[cf. Eq.~\eqref{eq:spectral_decomposition}] will benefit from an exponential 
OD speedup, in its approach to equilibrium at $T_{b}$. Instead, 
the system originally in equilibrium at $T_{c}$ will display a slower 
relaxation in the bath at $T_{b}$. This is regarded as the strong 
ME~\cite{klich2019}. Mind that, in general, $T_h$ will not be such that
$\mathbb{E}^{T_{h}}[\mathcal{M}^2_{w}]$ exactly equals
$\mathbb{E}^{T_{b}}[\mathcal{M}^2_{w}]$, but $T_h$ will 
be near the temperature where exact equality is achieved. That is, 
the condition \(0<|\alpha_2^{h}|<|\alpha_2^{c}|\) is fulfilled, 
which leads to a subexponential speedup regarded as the weak 
ME~\cite{klich2019}. 

In particular, the ME will be most spectacular if we choose $\mathcal{A}$ 
such that \(\mathbb{E}^{T_{c}}[\mathcal{A}]\) is closer to 
\(\mathbb{E}^{T_{b}}[\mathcal{A}]\) than 
\(\mathbb{E}^{T_{h}}[\mathcal{A}]\) is, because in that case the 
difference
\(\mathbb{E}_t^{h}[\mathcal{A}]-\mathbb{E}_t^{c}[\mathcal{A}]\)
will change sign in a clearer way.

\paragraph{Preheating for faster cooling.}
In these protocols~\cite{gal2020,pemartin2021} the transition of the system 
from equilibrium at temperature \(T_0\) toward equilibrium at bath 
temperature \(T_{b}<T_0\) can be done faster by introducing a brief 
sudden quench at a higher temperature \(T_{q}>T_0\), rather than simply 
leaving the system to freely relax under the action of the bath. 

In order to amplify the effect, we choose $T_0$ near the maximum of 
$\mathbb{E}^T[\mathcal{M}^2_{w}]$, so that 
$\alpha_2^{T_0\to T_{b}}$ will be as large as possible, cf. 
Eq.~\eqref{eq:astucia}. On the other hand, we choose $T_{q}\gg T_0$ 
so that 
$\mathbb{E}^{T_{b}}[\mathcal{M}^2_{w}]>
\mathbb{E}^{T_{q}}[\mathcal{M}^2_{w}]$, see 
Fig.~\ref{fig:anomalous_effects}(b). Now, starting from the equilibrated 
system at $T_0$, we instantaneously raise the bath temperature to 
$T_{q}$, and let the system relax. During the relaxation, the 
expected value of  $\mathcal{M}^2_{w}$ decreases from its initial 
value (which is the $T_0$ equilibrium value) and eventually crosses 
$\mathbb{E}^{T_{b}}[\mathcal{M}^2_{w}]$
at a time $t'$. This means [cf. Eq.~\eqref{eq:astucia}] that we can find a 
suitable time $t_w\approx t'$
such that, if  we instantaneously lower the bath temperature from 
$T_q$ to $T_{b}$ at time $t_w$, we shall start the relaxation
at $T_{b}$ 
with $\alpha_2^{t_w}=0$. The $t_w$ time overhead, due to the 
initial temperature excursion from $T_0$ to $T_{q}$, is compensated 
by the exponential speedup at $T_{b}$, that would otherwise be absent.

\paragraph{Heating and cooling may be asymmetric processes} (see  
also~\cite{lapolla2020,ibanez2023}). Let us consider the maximum of 
$\mathbb{E}^T[\mathcal{M}^2_{w}]$ at $T=T^*_b$ in 
Fig.~\ref{fig:anomalous_effects}(c). Given the aforementioned relation between 
metastability and slow relaxations, it is natural to expect that the largest
relaxation time in the system $1/|\lambda_2|$, see 
Eq.~\eqref{eq:spectral_decomposition}, will also attain its maximum value at
$T^*_{b}$ (the asymmetry stems from $\lambda_2$, rather than 
$\alpha_2$). Now, one would  naively think taking to equilibrium at 
$T_{A}$ a system initially at  equilibrium at $T_{B}$  would 
take the same time as the inverse process $T_{A} \to T_{B}$. 
Quite on the contrary, if we choose $T_{B}\!=\!T^*_{b}$  
the process $T_{B} \to T_{A}$ is \emph{faster} than 
its counterpart $T_{A} \to T_{B}$, no matter 
whether $T_{A} < T_{B}$ or $T_{A} > T_{B}$. 
Indeed, Eq.~\eqref{eq:astucia} tells us that 
$\alpha_2^{T_{B} \to T_{A}}$ and 
$\alpha_2^{T_{A} \to T_{B}}$ are numbers of similar magnitude 
(but opposite sign). Hence, the slowness of the relaxation is ruled by 
$1/|\lambda_2|$, which is larger at $T_B$.

The relaxation of the energy is an important exception, however. Indeed, 
applying Eq.~\eqref{eq:FDT} to \(\mathcal{M}^2_{w}\), one finds that
$\beta_2^{\mathcal{E}}$ [cf. Eq.~\eqref{eq:spectral_decomposition}] is 
inordinately small at $T_{B}\!=\!T^*_{b}$. Therefore, the 
approach to equilibrium of $\mathcal{E}$ at $T_{B}$ is ruled by 
$\lambda_3$ rather than $\lambda_2$, which precludes us from making 
definite predictions.

%%%%%%%%%%%%%%%%%%%%%%%%%%%%%%%%%%%%%%%%%%
%
%    SECTION: Ising model
%
%%%%%%%%%%%%%%%%%%%%%%%%%%%%%%%%%%%%%%%%%%
\paragraph{A working example: The antiferromagnetic 1D Ising model.}
We consider a periodic chain with \(N\) spins 
\(\sigma_i=\pm 1\), \(1\leq i\leq N\), and \(\sigma_{N+1}:=\sigma_1\). 
The state space is given by \(\Omega = \{-1,1\}^N\). 
The energy for a given spin configuration  
\(\boldsymbol{x} = (\sigma_1,\sigma_2,\ldots,\sigma_N)\) is 
\begin{equation}
\mathcal{E}(\boldsymbol{x}) \;:=\;
- J \sum\limits_{k=1}^N \sigma_k \, \sigma_{k+1}  -
h \sum\limits_{k=1}^N \sigma_k~,
\label{eq:def_E}
\end{equation}
where we assume $J<0$ and $h>0$, as well as  $N$ even to avoid frustration.  
The minimum energy configuration differs at both sides of the line $2J+h=0$. 
If $J>-h/2$ the ground state (GS) is the
uniform configuration $\{\sigma_i=1\}$. Instead, if $J<-h/2$ the GS is one 
of the two
AF ordered staggered configurations $\{\sigma_i=(-1)^i\}$ or 
$\{\sigma_i=(-1)^{i+1}\}$. Therefore, the first-order transition at 
$T\!=\!0$ needed to demonstrate exotic OD is realized in this model.

The uniform ($\mathcal{M}_u$) and the staggered 
($\mathcal{M}_\text{st}$) magnetizations are order parameters able to 
discriminate our GS:
\begin{equation}\label{eq:def-M}
 \mathcal{M}_u(\boldsymbol{x}) \;=\; \sum_{k=1}^{N}\sigma_k\,,\quad 
{\mathcal M}_\text{st}(\boldsymbol{x}) \;=\; \sum_{k=1}^{N}(-1)^k\sigma_k
\end{equation}
(for the uniform GS, $\mathcal{M}_u=N$ and $\mathcal{M}_\text{st}=0$, 
while for the staggered GSs one finds $\mathcal{M}_u=0$ and 
$\mathcal{M}_\text{st}=\pm N$).
The energy $\mathcal{E}$~\eqref{eq:def_E} is invariant under spatial 
translations ($\sigma_i\to\sigma_{i+1}$) which ensures that 
$\mathbb{E}^T[\mathcal{M}_\text{st}]=0$ for all temperatures. This is why 
we make stable the uniform GS by choosing \((J,h)=(-4,8.2)\). Hence, our 
\emph{wrong} order parameter
will be $\mathcal{M}_w\equiv \mathcal{M}_\text{st}$. 

Other magnitudes of interest will be
the staggered susceptibility $\chi_\text{st}$ and the spin-spin interaction
$\mathcal{C}_1$:
\begin{equation}\label{eq:def-chi_st-C1}
\chi_\text{st}\;=\;\frac{1}{N}\, \mathbb{E}^T[\mathcal{M}^2_\text{st}]\,,\quad 
\mathcal{C}_1(\boldsymbol{x}) \;=\; \sum_{k=1}^N\,\sigma_k \,\sigma_{k+1}\,.
\end{equation}
The reader will note that all our magnitudes of interest (namely, 
$\mathcal{M}_u, \, \mathcal{M}^2_\text{st},\,\mathcal{E}$, and 
$\mathcal{C}_1$) are invariant under spatial translations. Also our dynamics, 
see Eq.~\eqref{eq:master-equation}, preserves
the translation invariance of the starting probability $\bm{P}^{(t=0)}$.
Hence, the spectral decomposition~\eqref{eq:spectral_decomposition} can be 
restricted to the subspace of magnitudes $\mathcal{O}_k^b$ that are 
themselves invariant under translations.

Figure~\ref{fig:ansatz_check} shows that the two conditions necessary for 
exotic OD [namely, a nonmonotonic behavior of $\chi_\text{st}$ and a small
angle---as defined by Eq.~\eqref{eq:producto-escalar}---between 
$(\mathcal{M}_\text{st}^2)^\bot$ and $\mathcal{O}_2^{b}$] 
are met with our working parameters.

%%%%%%%%%%%%%%%%%%%%%%%%%%%%%%%%%%%%%%%%%%
%
%    SECTION: Results
%
%%%%%%%%%%%%%%%%%%%%%%%%%%%%%%%%%%%%%%%%%%

\paragraph{Results.} We have considered the three protocols explained in 
Fig.~\ref{fig:anomalous_effects} in the 1D AF Ising model considered above.
We have studied a single-site dynamics (heat-bath dynamics or Gibbs 
sampler~\cite{sokal1997functional,levin2017markov,supplemental}).
For short chains (\(N=8,12\)), we solved the master 
equation~\eqref{eq:master-equation} through
Monte Carlo (MC) simulations, and by finding the ``exact''
spectral decomposition of the operator \(R\)~\footnote{
   Strictly speaking, we computed the spectral decomposition of $R$ 
   using 300-digit arithmetic \cite{supplemental}.
}, 
full details can be found in~\cite{supplemental}.
The two methods were in full agreement and supported the proposed approach 
for small \(N=8, 12\). For larger chains (\(N=32\)), only MC simulations 
were computationally feasible and, again, validated our proposal. For 
clarity's sake, we only show numerical results for a selection observables 
(see~\cite{supplemental} for the remaining ones). In all three protocols,
we have found statistically compatible results for $N=12$ and $32$.

\begin{figure}[htb]
\includegraphics[width=8.6cm]{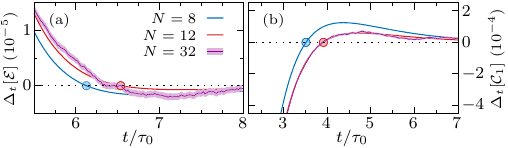}
\caption{Mpemba effect. Evolution of $\Delta_t[\mathcal{A}]$ [cf. 
Eq.~\eqref{eq:Delta_def}] for the observables $\mathcal{A}=\mathcal{E}$ (a), 
and $\mathcal{A}=\mathcal{C}_1$ (b). We show the results for $N=8$ (blue), 
$N=12$ (red), and $N=32$ (purple, with a lighter shade representing the error 
bars of the MC data). The time at which $\Delta_t[\mathcal{A}]$ changes sign is marked by a dot.}
\label{fig:mpemba_check}
\end{figure}

Figure~\ref{fig:mpemba_check} illustrates the ME as obtained with 
\(T_{b}=1\), \(T_c=4.15\), and \(T_h=15.177\), 
by a change of sign in 
\begin{equation}\label{eq:Delta_def}
\Delta_t[\mathcal{A}]\;:=\; \frac{1}{N}\, \left( 
\mathbb{E}_t^{h}[\mathcal{A}]-
\mathbb{E}_t^{c}[\mathcal{A}] \right) \,.
\end{equation}

\begin{figure}[htb]
\includegraphics[width=8.6cm]{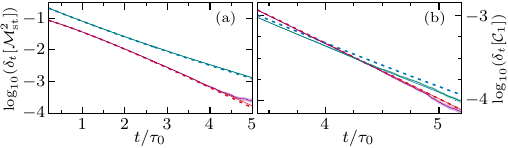}
\caption{Preheating strategy for faster cooling. Evolution of 
\(\delta_t[\mathcal{A}]\) [cf. Eq.~\eqref{eq:delta_def}] for 
$\mathcal{A}=\mathcal{M}_\text{st}^2$ (a),
and $\mathcal{A}=\mathcal{C}_1$ (b). Colors red and purple (respectively, 
blue and green) denote that the protocol includes (respectively does not 
include) an initial quench at temperature \(T_q\). We show data for 
$N=8$ (dashed lines), $N=12$ (solid lines), and $N=32$ (solid lines with a 
lighter shade representing the error bars of the MC data). Both panels 
show the expected speedup for the preheating protocol.}
\label{fig:supercooling_check}
\end{figure}

Supercooling through preheating is illustrated in 
Fig.~\ref{fig:supercooling_check}, with the choices \(T_{b}=1\), 
\(T_0=4.15\), \(T_q=2000\) and \(t_w=0.156\). It is useful to define
\begin{equation} \label{eq:delta_def}
\delta_t[\mathcal{A}] \;:=\; \frac{1}{N} \, \left| 
\mathbb{E}_t[\mathcal{A}] - \mathbb{E}^{T_b}[\mathcal{A}] \right| \,.
\end{equation}

\begin{figure}[htb]
\includegraphics[width=8.6cm]{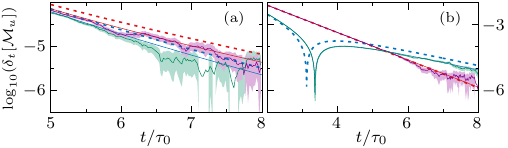}
\caption{(A)symmetry in heating and cooling. Evolution of 
\(\delta_t[\mathcal{M}_u]\) [cf. Eq.~\eqref{eq:delta_def}] for 
cooling (blue and green) and heating (red and purple) processes. We show 
data for $N=8$ (dashed lines), $N=12$ (solid lines), and $N=32$ (solid lines 
with a lighter shade representing the errors of the MC data). The cooling 
and heating processes were carried out between: (a) \(T_A=1\) and \(T_B=4.15\);
and (b) \(T_A=15.177\) and \(T_B=4.15\). Note that in panel (b), 
$\mathbb{E}_t[\mathcal{M}_{u}]-
\mathbb{E}^{T_{b}=T_B}[\mathcal{M}_{u}]$ changes sign at 
$t\approx 3$ for the system originally at $T_A=15.177$.
It is always slower to approach $T_B=4.15$: i.e. to heat up in (a), and to 
cool down in (b).}
\label{fig:asymmetry_check}
\end{figure}

Finally, the asymmetry between heating and cooling is illustrated in 
Fig.~\ref{fig:asymmetry_check}(a) with the choices $T_A=1$ and 
$T_B=4.15\approx T^*_b$,
and in Fig.~\ref{fig:asymmetry_check}(b) with the choices 
$T_A=15.177$ and $T_B=4.15$.

%%%%%%%%%%%%%%%%%%%%%%%%%%%%%%%%%%%%%%%%%%
%
%    SECTION: Discussion
%
%%%%%%%%%%%%%%%%%%%%%%%%%%%%%%%%%%%%%%%%%%
\paragraph{Discussion.} We have shown that the ``weak form'' of 
Markov dynamics provides a unified,
geometric framework that allows us to explain and control several exotic OD
effects pertaining to the Mpemba effect realm. Our approach departs from
previous work that usually privilege the ``strong form'' of the dynamics by
following the evolution of the system entropy (or, rather, some kind of
entropic ``distance'' such as the Kullback-Leibler divergence~\cite{KB51}). 
We have dealt,
instead, with different physical observables, some of which can be measured at
the mesoscopic level. Our geometric approach has unearthed an orthogonality
phenomenon that may cause an observable as prominent as the energy to remain
blind to the overall speedup achieved by the temperature changing protocols.
This result warns that one cannot have a too-narrow spectrum of observables
when investigating the Mpemba effect~\cite{Biswas23,Megias22}.

Finite-size effects on the separation of
timescales are also of concern, because it is this separation what 
determines the attainable exponential speedup. Fortunately, we have 
found that the speedup depends very mildly (if at all) on the system size. 
Finally, we stress that the approach presented here can also be applied 
to other systems where anomalous relaxation has been 
observed~\cite{ibanez2023,gal2020,kumar2020}, and where the nonmonotonicity 
of equilibrium thermodynamic observables is also present. 
Also, extending our approach to the analysis of the time-varying 
temperature protocols followed in recent
experiments~\cite{pires:23} is an exciting venue for future work.

\begin{acknowledgments}
We thank Oren Raz for discussions about the nonmonotonic behavior of the 
susceptibilities with temperature. 
We also thank Lamberto Rondoni for comments and for pointing out 
Refs.~\cite{evans20081,evans20082}.
This work was partially supported by Grants No.~PID2022-136374NB-C21, 
No.~PID2021-128970OA-I00, No.~PID2020-116567GB-C22, No.~FIS2017-84440-C2-2-P, 
and No.~MTM2017-84446-C2-2-R funded by Ministerio de Ciencia e Innovaci\'on
y Agencia Estatal de Investigaci\'on 10.13039/501100011033, by  
``ERDF A way of making Europe'', and by the European Union. A.L. was also
partly supported by FEDER/Junta de Andaluc\'{\i}a Consejer\'{\i}a de 
Universidad, Investigaci\'on e Innovaci\'on, and by the European
Union, through Grants No.~A-FQM-644-UGR20, and No.~C-EXP-251-UGR23.
J.S. was also
partly supported by the Madrid Government (Comunidad de
Madrid-Spain) under the Multiannual Agreement with UC3M in the line
of Excellence of University Professors (EPUC3M23), and in the
context of the V~PRICIT (Regional Programme of Research and
Technological Innovation). I.G.-A.P. was supported by the Ministerio de
Ciencia, Innovaci\'on y Universidades (MCIU, Spain) through FPU
Grant No.~FPU18/02665. We are also grateful for the computational
resources and assistance provided by PROTEUS, the supercomputing
center of the Institute Carlos~I for Theoretical and Computational
Physics at University of Granada, Spain.
\end{acknowledgments}

\bibliographystyle{apsrev4-2}
%\bibliography{ising1D.bib}

\end{document}

\title{Supplemental Material for ``Shortcuts of freely relaxing systems using 
equilibrium physical observables''}

\author{Isidoro Gonz\'alez-Adalid Pemart\'{\i}n}\affiliation{Departamento de 
F\'{\i}sica Te\'orica, Universidad Complutense, 28040 Madrid, Spain}%

\author{Emanuel Momp\'o}\affiliation{Departamento de Matem\'atica Aplicada, 
Grupo de Din\'amica No Lineal, Universidad Pontificia Comillas, 
Alberto Aguilera 25, 28015 Madrid, Spain}%
\affiliation{Instituto de Investigaci\'on Tecnol\'ogica (IIT), 
Universidad Pontificia Comillas, 28015 Madrid, Spain}%

\author{Antonio Lasanta}\email[Corresponding author: ]{alasanta@ugr.es}%
\affiliation{Departamento de \'Algebra, Facultad de Educaci\'on, Econom\'ia y 
Tecnolog\'ia de Ceuta, Universidad de Granada, Cortadura del Valle, s/n, 
51001 Ceuta, Spain}%
\affiliation{Nanoparticles Trapping Laboratory, Universidad de Granada, 
Granada, Spain}%

\author{V\'{\i}ctor Mart\'{\i}n-Mayor}\affiliation{Departamento de F\'{\i}sica 
Te\'orica, Universidad Complutense, 28040 Madrid, Spain}%
\affiliation{Instituto de Biocomputaci\'on y F\'{\i}sica de Sistemas Complejos 
(BIFI), 50018 Zaragoza, Spain}%

\author{Jes\'us Salas}
\affiliation{Departamento de Matem\'aticas, Universidad Carlos III de Madrid, 
28911 Legan\'es, Spain}%
\affiliation{Grupo de Teor\'{\i}as de Campos y F\'{\i}sica Estad\'{\i}stica, 
Instituto Gregorio Mill\'an, Universidad Carlos III de Madrid, Unidad Asociada 
al Instituto de Estructura de la Materia, CSIC, Spain}%

\date{\today}
\maketitle

In this Supplemental Material we will give extra details about our work. 
In Sec.~\ref{SMsect:Markov}, 
we will discuss the Markov dynamics, as well as the ways we have implemented it. 
In Sec.~\ref{SMsect:mean_values}, we will display exact formulae related 
to some expectation values 
of interest for a 1D Ising model in an external magnetic field. 
Finally, in Sec.~\ref{SMsect:Results}, we show some additional results 
that could not be displayed in the main text for lack of space.

\section{The Markov dynamics} \label{SMsect:Markov}
In order to explain our continuous-time dynamics, we shall follow this 
step-by-step approach:
\begin{enumerate}
\item We shall recall in Sec.~\ref{SMsect:discrete-time} the necessary
  mathematical ideas in the familiar context of discrete-time Markov Chain
  Monte Carlo (MC); for a more paused exposition see e.g.
  Refs.~\cite{sokal1997functional,levin2017markov}.
\item We show in Sec.~\ref{SMsect:lazy} how to obtain a lazy
  version of any discrete-time Markov chain, emphasizing how the basic
  formulae need to be modified for the lazy version of the 
  dynamics.~\footnote{%
     Our definition of lazy dynamics is a generalization of the definition 
     introduced in Sec.~1.3 of Ref.~\cite{levin2017markov}. 
}
\item The continuous-time Markov dynamics is straightforwardly obtained from
  the lazy discrete-time algorithm as explained in
  Sec.~\ref{SMsect:continuos}.
\item Finally, we recall in Sec.~\ref{SMsect:practical} how the $n$-fold way
  idea~\cite{bortz:75,gillespie:77} allows us to obtain a rejection-free
  algorithm for our Markov dynamics with continuous time.
\end{enumerate}

In order to give some flesh to the general formulae, we shall use the example
of the Ising spin chain that is studied in the main text. The inner product 
between
\emph{real} observables will also be defined here as in the main 
text~\footnote{For complex
  observables, one would modify Eq.~\eqref{SMeq:producto-escalar} as
  \protect{$\langle\,\mathcal{A}\, |\, \mathcal{B} \,\rangle =
    \sum_{\boldsymbol{x}\in\Omega}\, \pi_{\boldsymbol{x}}^{T_b}\,
    \overline{\mathcal{A}(\boldsymbol{x})}\, \mathcal{B}(\boldsymbol{x})\,,$}
  where the over-line stands for complex conjugation.
}:
\begin{equation}\label{SMeq:producto-escalar}
\langle\, \mathcal{A}\, |\, \mathcal{B} \,\rangle \;:=\; 
\sum_{\boldsymbol{x}\in\Omega}\, \pi_{\boldsymbol{x}}^{T_b}\, 
\mathcal{A}(\boldsymbol{x})\, \mathcal{B}(\boldsymbol{x})\,.
\end{equation}
Indeed, whether the dynamics are discrete or continuous time does not
make any difference in this respect.

Specifically, in examples we shall be referring to a periodic chain with \(N\)
spins \(\sigma_i=\pm 1\), \(1\leq i\leq N\). The state space is given by
\(\Omega = \{-1,1\}^N\). Hence, the number of states is $|\Omega|=2^N$.  
The energy for a given spin configuration
\(\boldsymbol{x} = (\sigma_1,\sigma_2,\ldots,\sigma_N)\) is
\begin{equation}\label{SMeq:Hamiltonian}
\mathcal{E}(\boldsymbol{x}) \;:=\;
- J \sum\limits_{k=1}^N \sigma_k \, \sigma_{k+1}  -
h \sum\limits_{k=1}^N \sigma_k~,
\end{equation}
where we assume \(\sigma_{N+1}:=\sigma_1\) due to the periodic boundary 
conditions. The chain length will be even, so there is no frustration 
in the antiferromagnetic regime (i.e., \(J<0\)).

The partition function of this model is given by 
\(Z_N(J,h;T_b) = \sum_{\boldsymbol{x}\in \Omega} 
\exp\left\{-\mathcal{E}(\boldsymbol{x})/(k_{B} T_{b})\right\}\), 
making
\begin{equation}\label{eq:pi_def}
\pi^{T_b}_{\boldsymbol{x}}(J,h) \;=\; 
\frac{\exp\left\{-\mathcal{E}(\boldsymbol{x})/(k_{B} 
T_{b})\right\}}{Z_N(J,h;T_{b})}
\end{equation}
to be the Boltzmann distribution \(\boldsymbol{\pi}^{T_b}\) for 
the couplings \((J,h)\) and the bath temperature \(T_{b}\).

The main examples of observables that we shall be considering are the energy
$\mathcal{E}$ \eqref{SMeq:Hamiltonian}, the uniform and staggered
magnetizations
\begin{equation}\label{SMeq:obs1}
\mathcal{M}_{u}(\boldsymbol{x}) \;=\; \sum_{k=1}^{N}\sigma_k\,,\quad 
\mathcal{M}_{\text{st}}(\boldsymbol{x}) \;=\; \sum_{k=1}^{N}(-1)^k\sigma_k\,,
\end{equation}
as well as the spin-spin interaction,
\begin{equation}\label{SMeq:c1}
\mathcal{C}_1(\boldsymbol{x}) \;=\; \sum_{k=1}^N\,\sigma_k \,\sigma_{k+1}\,.
\end{equation}
From these observables we compute the following expectation values: 
the energy density
$E = {\mathbb E}^T[\mathcal{E}]/N$, the uniform magnetization density
$M_u = {\mathbb E}^T[\mathcal{M}_u]/N$,
the spin-spin interaction
$C_1 = {\mathbb E}^T[\mathcal{C}_1]/N$, and the staggered susceptibility
$\chi_\text{st} = {\mathbb E}^T[\mathcal{M}_\text{st}^2]/N$. 
Note that the translation symmetry of the spin chain implies that 
${\mathbb E}^T[\mathcal{M}_\text{st}]=0$.
Besides, as in the main text, here $\bm{1}$ is the constant function
such that \(\boldsymbol{1}(\boldsymbol{x})=1\) for any state
$\boldsymbol{x}$.

\subsection{Discrete time}\label{SMsect:discrete-time}
Let \(T_{\bm{x},\bm{y}}\) be the probability for the system to
jump from state \(\bm{x}\) to state \(\bm{y}\) in a single
time-step. If the number of states $|\Omega|$ is finite, 
\(T_{\bm{x},\bm{y}}\) can be regarded as an element
of an $|\Omega|\times|\Omega|$ matrix $T$.

The Markov and stationary properties of the dynamics are encoded in the 
so-called master
equation. Let us denote $P^{(m)}_{\bm{x}}$ the probability of finding the
  system in state $\bm{x}$ at the discrete time-step $m$. Then, the
  probability after $n$ further time-steps is
\begin{equation}\label{SMeq:Master-eq}
  P^{(m+n)}_{\bm y}\;=\;\sum_{\bm{x}\in\Omega} P^{(m)}_{\bm{x}}
  [T^n]_{\bm{x},\bm{y}}\,,
\end{equation}
where $T^n$ is the $n$-th power of matrix $T$. In particular, we have the
time evolution of the probability from the initial condition 
$P^{(n=0)}_{\bm{x}}$:
\begin{equation}\label{SMeq:discrete-time-evolution}
  P^{(n)}_{\bm y}\;=\;\sum_{\bm{x}\in\Omega} P^{(n=0)}_{\bm{x}}
  [T^n]_{\bm{x},\bm{y}}\,,
\end{equation}
Mind that, in this formalism, probabilities are regarded as row vectors that
are right-multiplied by matrix $T$ to get a new probability (hence, a new
row vector). 

The main properties that $T$ should
fulfill are positivity and completeness of the conditional probabilities,
\begin{equation}\label{SMeq:positive-complete}
T_{\boldsymbol{x},\boldsymbol{y}} \geq 0\,, \, \forall 
\bm{x},\bm{y}\in\Omega\,;\quad 
\sum_{\bm{y}\in\Omega}\, T_{\bm{x},\bm{y}}\;=\; 1\,, \, \forall 
\bm{x}\in\Omega\,,
\end{equation}
stationary,
\begin{equation}\label{SMeq:balance}
\pi_{\boldsymbol{y}}^{T_b} \;=\; \sum_{\bm{y}\in\Omega}\,
  \pi_{\boldsymbol{x}}^{T_b} \, T_{\bm{x},\bm{y}}\,,
\end{equation}
and irreducibility,
\begin{equation}
  \forall \bm{x},\bm{y}\in\Omega\,, \exists\, n_{\bm{x},\bm{y}}>0 \text{ such
    that } [T^{n_{\bm{x},\bm{y}}}]_{\bm{x},\bm{y}}\;>\;0\,.
\end{equation}
Note that the stationary condition implies that the Boltzmann weight
$\bm{\pi}^{T_b}$ is a left eigenvector of the matrix $T$, so that a 
system
initially in thermal equilibrium at temperature $T_{b}$---i.e., 
distributed according to $\bm{\pi}^{T_b}$---remains 
in equilibrium forever, see
Eq.~\eqref{SMeq:discrete-time-evolution}. Irreducibility means that
any state of the phase space $\Omega$ should be eventually reachable from 
any starting state $\bm{x}$.

Just to make one example, let us consider the Ising spin chain. The
corresponding heat-bath discrete-time dynamics with random-access to the chain
is encoded in the following way. If two, or more, spins take different 
values in configurations \(\boldsymbol{x}\) and \(\boldsymbol{y}\), then
$T_{\bm{x},\bm{y}}=0$. This is why this dynamic is sometimes described as
single spin-flip. If $\bm{x}$ and $\bm{y}$ differ in the value of just one
spin, then
\begin{eqnarray}\label{SMeq:heat-bath}
T_{\bm{x},\bm{y}}&=&\frac{1}{N} \, R^\text{HB}_{\bm{x},\bm{y}}\,,\\
R^\text{HB}_{\bm{x},\bm{y}}&=&\frac{\exp\left\{-[\mathcal{E}(\boldsymbol{y})-
\mathcal{E}(\boldsymbol{x})]/(k_B T_{b})\right\}}%
{1+\exp\left\{-[\mathcal{E}(\boldsymbol{y})-
\mathcal{E}(\boldsymbol{x})]/(k_B T_{b})\right\}}\,.
\label{SMeq:heat-bath-accept}
\end{eqnarray}
[Diagonal matrix elements $T_{\bm{x},\bm{x}}$ are fixed by the completeness
condition~\eqref{SMeq:positive-complete}]. Mind that matrix $T$ is very
sparse: every row contains $2^{N}$ matrix elements, but only $N+1$ of them are
different from zero (i.e., the diagonal term, and the $N$ states $\bm{y}$ that 
are connected with $\bm{x}$ by a single spin flip). In fact, 
the $1/N$ prefactor in Eq.~\eqref{SMeq:heat-bath} tells us that the spin 
that will be attempted to 
flip is chosen with uniform probability. The probability of accepting the
spin-flip $R^\text{HB}_{\bm{x},\bm{y}}$ ensures that $T$ verifies the detailed 
balance condition
\begin{equation}\label{eq:SM-detailed-balance}
    \pi_{\boldsymbol{x}}^{T_b}\, T_{\bm{x},\bm{y}} \;=\;
    \pi_{\boldsymbol{y}}^{T_b}\, T_{\bm{y},\bm{x}}\,,
\end{equation}
that can be straightforwardly combined with the completeness condition to
show that detailed balance implies
stationary~\eqref{SMeq:balance}~\footnote{The reverse statement is not true:
stationary does not entail detailed balance.}. Showing that heat-bath
verifies the other conditions (positivity, completeness, and irreducibility 
for positive temperature) is a textbook exercise.  
  
Note as well that for single spin-flip dynamics it
is customary to restrict the $n$ in Eq.~\eqref{SMeq:discrete-time-evolution} to
a multiple of $N$,
\begin{equation}\label{SMeq:extensive-change-attempts}
n\;=\;k\, N\,,\quad  k \;\in\; \mathbb{N}\,,
\end{equation}
so that every spin has (on average) $k$ opportunities to be flipped.

At this point, it would be natural to solve
Eq.~\eqref{SMeq:discrete-time-evolution} by finding a basis of
left eigenvectors. However, it will prove useful to diagonalize instead the
related operator ${\cal T}$~\footnote{For any square matrix, it is possible to 
  find a basis of the vector space formed solely by left eigenvectors if, 
  and only if, it is also possible to find a basis of right eigenvectors. 
  Furthermore, the spectrum 
  of left eigenvalues is identical to the spectrum of right eigenvalues.
}.   
Given an observable $\mathcal{A}$, 
we obtain a new observable ${\cal T}[\mathcal{A}]$ as
\begin{equation}
  {\cal T}[\mathcal{A}](\bm{x}) \;=\; \sum_{\bm{y}\in\Omega}
  T_{\bm{x},\bm{y}}\, \mathcal{A}(\bm{y})\,.
\end{equation}
So, in this formalism, observables are identified with column vectors that
are left-multiplied by the matrix $T$ to get a new column vector, hence a new
observable. The interpretation of the new observable ${\cal T}^n[\mathcal{A}]$
is simple: take a system initially in state $\bm{x}$, then 
${\cal T}^n[\mathcal{A}](\bm x)$ is the expected value of
$\mathcal{A}$ after $n$ dynamical steps~\footnote{
   Therefore, the approach to equilibrium means that ${\cal T}^n[\mathcal{A}]$ 
   approaches
   $\mathbb{E}^{T_b}[\mathcal{A}]\bm{1}$ as $n$ grows. In other 
   words, after many time steps, the expected value for $\mathcal{A}$ becomes 
   the equilibrium expectation value at temperature $T_{b}$, 
   irrespective of
   the starting configuration $\bm{x}$
}. The completeness condition implies as
well that ${\cal T}[\bm{1}]=\bm{1}$ (i.e., $\bm{1}$ is a right eigenvector of
$\cal{T}$ with eigenvalue one).

Now, the crucial observation is that, as the reader can easily show, detailed
balance implies that ${\cal T}$ is a self-adjoint operator for the
inner-product~\eqref{SMeq:producto-escalar},
\begin{equation}
  \langle\, {\cal T}[\mathcal{A}] \,|\,\mathcal{B} \, \rangle \;=\;
  \langle\, \mathcal{A} \,|\,{\cal T}[\mathcal{B}] \,\rangle\,.
\end{equation}
It follows that the spectrum of ${\cal T}$ (and hence of $T$) is
real. In addition, it follows from the completeness
condition~\eqref{SMeq:positive-complete} that all eigenvalues belong to the 
interval $[-1,1]$ (hence, the eigenvalue for the
constant functions, $\Lambda_1=1$, is the largest one).
Furthermore,
one can find an orthogonal basis of right eigenvectors---orthogonal with 
respect to the inner
product~\eqref{SMeq:producto-escalar}---\((\boldsymbol{1},
\mathcal{O}_2^{b},\mathcal{O}_3^{b},\ldots)\):
\begin{eqnarray}
{\cal T}[\mathcal{O}_k^{b}] &=&\Lambda_k\, 
\mathcal{O}_k^{b}\,,\\
1\;=\;\Lambda_1&\geq &\Lambda_2\;\geq\;\Lambda_3\;\geq\;\ldots\;\geq\;
\Lambda_{|\Omega|}\;\geq\; -1\,.\label{SMEq:eigenvalue-1}
\end{eqnarray}
One has a corresponding basis of left eigenvectors in which the starting
probability can be linearly expressed ($\bm{v}_1=\bm{\pi}^{T_b}$, 
of course)
\begin{equation}
\bm{v}_k \, T\;=\;\Lambda_k \, \bm{v}_k\,,\quad \bm{P}^{(n=0)}\;=\;
\bm{\pi}^{T_b}+\sum_{k=2}^{|\Omega|}
\gamma_k\, \bm{v}_k\,.
\end{equation}
Hence, the dynamic evolution for the probability is
\begin{equation}\label{SMeq:DD1}
  \bm{P}^{(n)}\;=\;\bm{\pi}^{T_{b}}+\sum_{k\ge2}\, \Lambda_k^n \, 
             \gamma_k\,  \bm{v}_k\,,
\end{equation}
while the discrete-time evolution for the expectation value
of an arbitrary (finite-variance) magnitude $\mathcal{A}$ is
\begin{eqnarray}\label{SMeq:DD2}
  \mathbb{E}_{n}[\mathcal{A}] &=& \mathbb{E}^{T_b}[\mathcal{A}] + 
\sum_{k\geq2} \alpha_k^{(t=0)}\, \beta_{k}^{\mathcal{A}}\, \Lambda_k^n\,,\\
  \label{SMeq:DD3a}
  \beta_{k}^{\mathcal{A}}&=&\langle O_k^{b}\,|\,\mathcal{A}\rangle\,, 
\\[2mm]
  \alpha_k^{(t=0)} &=& \sum_{\boldsymbol{x}\in\Omega}\,
  P^{(n=0)}_{\bm{x}}\,{\cal O}_k^{b}(\boldsymbol{x})\,. \quad\quad 
  \label{SMeq:DD3b}     
\end{eqnarray}
             
\subsection{The lazy discrete-time dynamics}\label{SMsect:lazy}
Let us now assume that we modify our discrete-time dynamics in the following
way. At each time step, with probability $\epsilon$ we attempt to modify the
system exactly as explained in Sec.~\ref{SMsect:discrete-time}, while with
probability $1-\epsilon$ we do nothing. The rationale for this modification is
the following: We want $\epsilon$ to represent a very small time step
(remember that our final goal is formulating a continuous time dynamics). 
Of course, in a
very short time interval it is highly unlikely that the system changes.

Mathematically, the matrix $T_\epsilon$ that represents the lazy dynamics 
can be simply written in terms of the identity matrix $\mathbb{I}$ and the 
matrix $T$ considered in Sec.~\ref{SMsect:discrete-time}
\begin{equation}\label{SMeq:epsilon-def}
  T_\epsilon\;=\;(1-\epsilon)\, \mathbb{I} + \epsilon\, T\,,
\end{equation}
or, better,
\begin{equation}\label{SMeq:R-def}
T_\epsilon \;=\; \mathbb{I}+\epsilon \, \widetilde R\,,\quad \widetilde R\;=\;
T-\mathbb{I}\,.
\end{equation}
Mind that the off-diagonal elements of $T$ and $\widetilde R$ are identical 
while, in terms of $\widetilde R$, the completeness 
relation~\eqref{SMeq:positive-complete} reads
\begin{equation}\label{SMeq:complete-2}
  \widetilde R_{\bm{x},\bm{x}}= -\sum_{\bm{y}\in\Omega\setminus \{\bm{x}\}}\,
\widetilde R_{\bm{x},\bm{y}}\,, \quad  \forall\bm{x}\in\Omega\,.
\end{equation}
It is also crucial that $\widetilde R$ and $T$ share the orthogonal basis of 
right eigenvectors
\((\boldsymbol{1},\mathcal{O}_2^{b},\mathcal{O}_3^{b},\ldots)\),
as well as the basis of left eigenvectors
\((\bm{\pi}^{T_b},\bm{v}_2,\bm{v}_3,\ldots)\). 

Thus, Eqs.~\eqref{SMeq:DD1}--\eqref{SMeq:DD3b} carry over to the case
of the lazy dynamics with the only change that the eigenvalues
$\Lambda_k$ of the matrix $T$ need to be replaced  by the eigenvalues
$\Lambda_{k,\epsilon}$ of the matrix $T_\epsilon$. 
Both sets of eigenvalues can be
expressed in terms of the eigenvalues $\widetilde\lambda_k$ of the matrix 
$\widetilde R$ (remember that
these three matrices, $T$, $T_\epsilon$ and $\widetilde R$ share both bases of 
left and right eigenvectors):
\begin{eqnarray}
  \widetilde \lambda_k&=&\Lambda_k -1\,,\\
  \Lambda_{k,\epsilon}&=&1+\epsilon\, \widetilde\lambda_k\,,
\end{eqnarray}
where
\begin{equation}
0\;=\;\widetilde\lambda_1\;\geq\;\widetilde\lambda_2\;\geq\;\ldots\;\geq\;
\widetilde\lambda_{|\Omega|}\;\geq\; -2\,.\\
\end{equation}
Hence, for $\epsilon<1/2$, all the eigenvalues of $T_\epsilon$ are guaranteed 
to be positive.

\subsection{The continuous-time dynamics}\label{SMsect:continuos}
We shall reach the limit of continuous time by making the parameter $\epsilon$
in Eq.~\eqref{SMeq:epsilon-def} arbitrarily small.  Let us start by considering that we need to reach a time $t$ such that
 $t/\tau_0$ is a rational number ($\tau_0$ is our time unit)
\begin{equation}\label{eq:fraccion}
\frac{t}{\tau_0}\;=\;\frac{p}{q} \;\in\; \mathbb{Q}\,, 
\end{equation}
where $p/q$ is an irreducible fraction (the case of irrational values of 
$t/\tau_0$ will be trivially solved by continuity through our final formulae). 
Next, we set a sequence of $\epsilon_r$ going to zero as
\begin{equation}
  \epsilon_r \;=\; \frac{1}{q r}\,,\quad r \;\in\; \mathbb{N}\,.
\end{equation}
So, fixing the number of time steps $n_r$ as
\begin{equation}
  n_r\;=\;N p r\,,   
\end{equation}
we find that $n_r\epsilon_r=N t/\tau_{0}$, irrespective of $r$.
The rationale for introducing the factor of $N$ is that we are already
planning to work with a single spin-flip dynamics, 
hence we wish to work with an extensive
number of spin-flip attempts [cf. Eq.~\eqref{SMeq:extensive-change-attempts}].

Eqs.~\eqref{SMeq:DD1} and~\eqref{SMeq:DD2} now take the form [the numerical 
coefficients $\beta_{k}^{\mathcal{A}}$ and $\alpha_k^{(t=0)}$
are the same of Eq.~\eqref{SMeq:DD3a}/\eqref{SMeq:DD3b}, respectively]
\begin{eqnarray}\label{SMeq:DD1bis}
\bm{P}^{(n_r)} &=& \bm{\pi}^{T_b}+\sum_{k\ge 2}\, 
(1+\epsilon_r\widetilde\lambda_k)^{n_r} \, 
 \gamma_k \, \bm{v}_k\,, \\[2mm]
\label{SMeq:DD2bis}
  \mathbb{E}_{n_r}[\mathcal{A}] &=& \mathbb{E}^{T_b}[\mathcal{A}] +
  \sum_{k\geq2} \alpha_k^{(t=0)} \, \beta_{k}^{\mathcal{A}}\, 
  (1+\epsilon_r\widetilde\lambda_r)^{n_r}\,. \qquad
\end{eqnarray}
Now, the limit of continuous time is reached by letting $r$
go to infinity, so that $\epsilon_r$ goes to zero:
\begin{equation}
  \lim_{r\to\infty}
  (1+\epsilon_r\widetilde\lambda_r)^{n_r}=
\mathrm{e}^{N\widetilde\lambda_k t/\tau_0}\,.
\end{equation}
(Recall that the eigenvalues $\widetilde{\lambda}_k$ are nonpositive.)
Our final expressions follow:
\begin{eqnarray}\label{SMeq:DD1tris}
  \bm{P}^{(n_r)} &=& \bm{\pi}^{T_b}+\sum_{k\ge 2}\ 
\mathrm{e}^{-N|\widetilde\lambda_k| t/\tau_0} 
           \gamma_k \bm{v}_k\,, \\[2mm]
            \label{SMeq:DD2tris}
  \mathbb{E}_{n_r}[\mathcal{A}] &=& \mathbb{E}^{T_b}[\mathcal{A}] +
  \sum_{k\geq2} \alpha_k^{(t=0)} \beta_{k}^{\mathcal{A}}\, 
 \mathrm{e}^{-N|\widetilde\lambda_k| t/\tau_0}\,.
\end{eqnarray}
In order to make contact with the expressions in the main text, we just need
to redefine our matrix $\widetilde R$ and the corresponding eigenvalues as
\begin{eqnarray}
  R^\text{HB} &=&N\, \widetilde R\,,\label{SMeq:normalizacion}\\
  \lambda_k    &=& N\, \widetilde\lambda_k\,.
\end{eqnarray}
      
\subsection{Practical recipes and description of our computations}
\label{SMsect:practical}
Given that in our Ising spin chain $|\Omega|=2^N$, we can carry out a fully
analytical computation only for moderate values of $N$, say  $N\leq 12$. For 
larger values of $N$, we have turned to a MC method. Let us explain the 
two approaches separately. For simplicity, we will denote hereafter the 
matrix $R^\text{HB}$ as $R$.

\subsubsection{Exact diagonalization}

It is crucial to observe from Eqs.~\eqref{SMeq:R-def} 
and~\eqref{SMeq:normalizacion}
that the matrix $R$ does not depend on $\epsilon$ in any way. Hence, we can 
obtain $R$ directly in the limit of a continuous-time dynamics.

For an Ising chain of length $N$, $R$ is a $2^N\times 2^N$ square matrix.
Indeed, it is a very
sparse matrix because only $N+1$ elements in each row are non vanishing.
The $N$ nonvanishing off-diagonal elements are those where the initial 
configuration $\bm{x}$ and the final one $\bm{y}$ differ by a single spin-flip.
These nonvanishing off-diagonal
matrix elements are identical to the probabilities for accepting the 
corresponding spin-flip in
Eq.~\eqref{SMeq:heat-bath-accept}.  The diagonal
elements are given, instead, by the completeness relation:
\begin{equation}
R_{\boldsymbol{x},\boldsymbol{x}}\;=\;-\sum_{\boldsymbol{y}\in
\Omega \setminus \{\bm{x}\}}
R_{\boldsymbol{x},\boldsymbol{y}}\,,\quad \forall \bm{x}\in\Omega.    
\end{equation}

We compute the eigenvalues and right eigenvectors of the matrix $R$ using
{\sc Mathematica}. In the first step, we calculated the matrix $R$ and 
the probability 
density $\bm{\pi}^{T_b}$ \emph{symbolically} as 
functions of the couplings $(J,h)$, and the bath temperature $T_b$. 
We then
checked \emph{symbolically} their basic properties: detailed balance, 
$R\, \bm{1} = \bm{0}$, and $\bm{\pi}^{T_b}\, R = \bm{0}$,
where $\bm{0}$ is the constant zero function; i.e., $\bm{0}({\bm{x}}) = 0$ 
for all $\bm{x}\in\Omega$.

The second step consists in evaluating the symbolic expressions for $R$ 
and $\bm{\pi}^{T_b}$
with high precision (i.e., 300-digit arithmetic). We then compute the 
eigenvalues $(0,\lambda_2,\lambda_3,\ldots)$
and right eigenvectors 
$(\bm{1},\mathcal{O}_2^{b},\mathcal{O}_3^{b},\ldots)$ of $R$. 
This is achieved by following Ref.~\cite[proof of Lemma 12.2]{levin2017markov}: if $D$ is the diagonal matrix 
whose diagonal elements are 
$D_{\bm{x},\bm{x}} = (\pi_{\bm{x}}^{T_b})^{1/2}$, then the matrix 
$R$ is similar to the symmetric (and real) matrix $A = D\, R \, D^{-1}$. 
The spectral theorem for this kind
of matrices ensures the existence of an \emph{orthonormal} basis of right 
eigenvectors 
$\{ \mathcal{P}_k\}_{k=1}^{|\Omega|}$ with respect to the inner product 
\begin{equation}
\langle\, \mathcal{A} \mid \mathcal{B} \,\rangle_0 \;=\; 
\sum\limits_{\bm{x}\in\Omega} \mathcal{A}(\bm{x})\, \mathcal{B}(\bm{x})\,,
\end{equation}
and such that the eigenvector $\mathcal{P}_k$ corresponds to the eigenvalue 
$\lambda_k$. This is so, because if we define 
$\mathcal{O}_k^{b} = D^{-1}\, \mathcal{P}_k$,
then $\mathcal{O}_k^{b}$ is a right eigenvector of $R$ with eigenvalue 
$\lambda_k$.
In this way we obtained the sets $\{\mathcal{O}_k^{b}\}_{k=1}^{|\Omega|}$ 
and
$\{\lambda_k\}_{k=1}^{|\Omega|}$. Indeed, we verified to high accuracy 
(e.g., at least 270 digits) 
that these eigenfunctions form an orthonormal basis with respect to the 
inner product \eqref{SMeq:producto-escalar}, and they satisfied that 
$R\,\mathcal{O}_k^{b} = \lambda_k\, \mathcal{O}_k^{b}$ for all $k$.

The final step needs the initial temperature $T_0$ of the process. For this 
temperature, we compute the initial 
probability density $\boldsymbol{P}^{(t=0)}$ as the Boltzmann weight at 
temperature $T_0$. Then we can compute the numerical constants 
$\alpha^{(t=0)}_k$ \eqref{SMeq:DD3b}.
Moreover, for all the observables $\mathcal{A}$ we want to consider, we also 
compute the constants $\beta_{k}^{\mathcal{A}}$
\eqref{SMeq:DD3a}. Now we can obtain the evolution of the expected value 
of $\mathcal{A}$ with time [cf. \eqref{SMeq:DD2tris}]
\begin{equation} \label{SQMec::EE1}
  \mathbb{E}_{t}[\mathcal{A}] \;=\; \mathbb{E}^{T_{b}}[\mathcal{A}] +
  \sum_{k\geq2} \alpha_k^{(t=0)} \beta_{k}^{\mathcal{A}}\, 
\mathrm{e}^{-|\lambda_k| t}\,,
\end{equation}  
where we have chosen units such that $k_B=\tau_0=1$. With this choice, 
the time is in units of a lattice sweep. 
Using this procedure, we have been able to deal with systems of length 
$N\le 12$; notice that for $N=12$, $\text{dim}(R)=4\,096$. 
For the next system length $N=14$, $\text{dim}(R)=16\,384$, which is beyond 
our computing capabilities.

\subsubsection{The Monte Carlo Algorithm}
In order to obtain a workable MC method one just needs
to go back to Sec.~\ref{SMsect:continuos}, and compute the probability of
not making any change whatsoever to configuration $\bm{x}$ in a time interval
$t$. It is straightforward to show that
\begin{equation}
  \lim_{r\to\infty} [T_{\epsilon_r}^{n_r}]_{\bm{x},\bm{x}}\;=\;
\mathrm{e}^{-|R_{\bm{x},\bm{x}}| t/\tau_0}\,.
\end{equation}
Hence, the probability  for the time of the first change to the configuration 
is the one of a Poisson
process, and we are in the canonical situation for an $n$-fold way
simulation~\cite{bortz:75,gillespie:77}. For the reader convenience, let us
briefly recall how one such simulation is carried-out. 

Let us imagine that we
want to update our system for times $t\in[t_1,t_2]$ starting from the 
configuration $\bm{x}_\text{ini}$ at time $t_1$. So, we first set our
clock to $t=t_1$, and carry out the following procedure (we name $\bm{x}$ 
the current spin configuration):
\begin{enumerate}
\item Select a time increment
  \begin{equation}
    \Delta t \;=\; \frac{\tau_0}{|R_{\bm{x},\bm{x}}|}\, \log(u)\,
  \end{equation}
  where $u$ is a uniformly distributed random number in the unit 
  interval $(0,1]$.
\item If $t+\Delta t >t_2$, stop the simulation [if some
  observable is to be computed for a time $t'\in(t,t_2)$, it can be computed 
  on the final spin configuration]. If $t+\Delta t <t_2$, go to (3). 
\item Update the clock of the simulation $t\to t+\Delta t$. Note
  that the spin configuration is constant for all times $t'\in (t,t+\Delta
  t)$. So, if we are to compute an observable for such a $t'$, it should be
  computed \emph{before} the spin configuration changes.
  
\item Choose the spin to be flipped with probability  
  $R_{\bm{x},\bm{y}}/|R_{\bm{x},\bm{x}}|$ (mind that if $t+\Delta t>t_2$ 
  this step should be skipped). Remember that $R_{\bm{x},\bm{y}}\neq 0$ only 
  if the new state $\bm{y}$ differs from the old state $\bm{x}$ precisely 
  by one spin-flip. Set the chosen state $\bm{y}$ as the new current state.
  
\item Return to (1).
\end{enumerate}
For an Ising chain, there is a simple technical trick to considerably speed up
the simulation. Indeed, Eq.~\eqref{SMeq:heat-bath-accept} tells us that the
$R_{\bm{x},\bm{y}}$ depends only on the energy change when flipping one of the
spins in the starting configuration $\bm{x}$. Now, the energy-change upon
flipping the spin $\sigma_k$ depends solely on $\sigma_{k-1}+\sigma_{k+1}$ 
(three
possible values: $2,0,-2$), and on the value of $\sigma_k=\pm 1$. Hence,
according to the energy change caused by their spin flip, there are only
$3\times 2$ possible kind of spins. Therefore, in order to compute $\Delta t$
and to decide efficiently which spin to update, it suffices to maintain an
updated set of six list of spins (one for every kind of spins).

\subsubsection{Our simulations}

We generate a large number of statistically independent trajectories $S$. 
On each
trajectory we consider observables $\mathcal{A}$ [see, for instance, 
Eqs.~\eqref{SMeq:obs1} and~\eqref{SMeq:c1}]. We also select a priori a mesh of
measuring times $\{t^{(1)},\ldots,t^{(M)}\}$. For each of this times, 
$t^{(k)}$ and for each trajectory $s$, we obtain the corresponding value 
\begin{equation}
\mathcal{A}_{k,s}\;\equiv\;\mathcal{A}[ \bm{x}^{(s)}(t^{(k)})]
\end{equation}
where $\bm{x}^{(s)}(t^{(k)})$ is the configuration of the $s$-th trajectory 
at time $t^{(k)}$. The expected value  $\mathbb{E}_{t^{(k)}}[\mathcal{A}]$ 
is estimated as
\begin{equation}
\mathbb{E}_{t^{(k)}}[\mathcal{A}] \;=\; \frac{1}{S}\sum_{s=1}^S 
\mathcal{A}_{k,s}\,,
\end{equation}
while we compute errors for these expected values in the standard way:
\begin{equation}
\Delta \mathbb{E}_{t^{(k)}}[\mathcal{A}] \;=\; \sqrt{\frac{1}{S(S-1)} 
\sum_{s=1}^S \Big(\mathcal{A}_{k,s} -
\mathbb{E}_{t^{(k)}}[\mathcal{A}]\Big)^2}\,.
\end{equation}

All our trajectories contain a preparation step and a measuring step. In the 
preparation step, the initial configurations for the trajectories are chosen 
randomly with uniform probability. Hence, the dynamics sketched in the 
previous paragraph is followed at the initial temperature for a time 
long enough to ensure
equilibration. Then, one (or more) temperature jumps are carried out. This 
amounts to changing the acceptance in Eq.~\eqref{SMeq:heat-bath-accept}, 
while taking as
the initial configuration at the new temperature, the final configuration 
at the previous temperature.

Technical simulation details are shown in Table \ref{tab:simul}. 
The simulations were performed in PROTEUS cluster, the supercomputing
center of the Institute Carlos~I for Theoretical and Computational
Physics at University of Granada. In particular, for each replica nodes of 
12 processors at $3.45$GHz and $16$Mb of memory were used. The parameters of 
the simulations were $J=-4$, $h=8.2$, and we choose time units such that 
$\tau_0=1$. The values presented in the main manuscript were obtained 
averaging a total number of $2.4 \times 10^9$ trajectories.

\begin{table}[t]
\begin{center}
\begin{tabular}{| c | c | c | c | c |}
\hline
Simulation $T_b, (T_q)$ & Size & Trajectories & Replicas & Time
\\ \hline
$15.177 \to 1$ & 8,12,32 & $2 \times 10^7$& 120 & 10 \\
$4.15 \to 1$ & 8,12,32 & $2 \times 10^7$ & 120 & 10 \\
$1 \to 15.177$ & 8,12,32 & $2 \times 10^7$ & 120 &10 \\
$4.15 \to 15.177$ & 8,12,32 & $2 \times 10^7$ & 120 & 10 \\ 
$15.177 \to 4.15 $ & 8,12,32 & $2 \times 10^7$ & 120 & 8 \\ 
$ 1 \to 4.15$ & 8,12,32 & $2 \times 10^7$ & 120 & 8 \\ 
$ 4.15 \to (2000) \to 1 $ & 8,12,32 & $2 \times 10^7$ & 120 & 10 (0.156)\\\hline
\end{tabular}
\caption{List of simulations. Columns: (I) Bath temperatures used in each 
case simulated. Parentheses indicate the temperature used during the 
first step in the preheating protocol. (II) Number of spins in the 
chain. (III) Number of independent trajectories simulated per replica. 
(IV) Independent simulated copies of the system. (V) Time length in 
units such that $\tau_0=1$. Parentheses indicate the time in contact 
with the first thermal bath during the preheating protocol.}
\label{tab:simul}
\end{center}
\end{table}

\section{Exact expectation values for some quantities} 
\label{SMsect:mean_values}

In order to present these expectation values, we first need to introduce some 
notation. Let us define the couplings 
\begin{eqnarray}
K \;=\; \frac{J}{k_B\, T}\,, \qquad H \;=\; \frac{h}{k_B\, T}\,.
\end{eqnarray}
All the following mean values depend on both quantities and on the spin 
chain length $N$ (recall that in our units, $k_B=1$). 

Basically, we will use the transfer-matrix method in this computation. 
We first define the
coefficient $\psi = \psi(K,H)$ as the ratio of the two eigenvalues 
of the transfer matrix:
\begin{equation}
\psi(K,H) \;=\; \frac{\cosh(H) - \sqrt{e^{-4K} + \sinh^2(H)}}
                     {\cosh(H) + \sqrt{e^{-4K} + \sinh^2(H)}} \,. 
\end{equation}
Notice that $|\psi(K,H)|< 1$, so $\lim_{N\to\infty}\psi^N = 0$. 

The uniform magnetization is given by
\begin{equation}
M_u \;=\; \frac{ \sinh(H) }{\sqrt{ e^{-4K} + \sinh^2(H)}} \, 
\frac{1 - \psi^N}{1 + \psi^N}\,.
\end{equation}
The spin-spin interaction is given by 
\begin{multline}
  C_1 \;=\; \frac{\sinh^2(H)}{e^{-4K} + \sinh^2(H)} \\[2mm] + 
 \frac{1+\psi^{N-2}}{1 + \psi^N}\, \frac{\psi}{1 + e^{4K}\,\sinh^2(H)} \,. 
\end{multline}
The internal energy is given by a linear combination of the previous two values
\begin{equation}
    E \;=\; -J \, C_1 - h \, M_{u} \,.
\end{equation}
The staggered susceptibility is given for a spin chain of even length by
\begin{equation}
\chi_\text{st} \;=\; \frac{e^{-4K}}{ \cosh(H)\, \sqrt{e^{-4K} + \sinh^2(H)}}\, 
\frac{1-2\psi^N}{1+\psi^N} \,.
\end{equation}

\section{Some additional results for the 1D Ising model} \label{SMsect:Results}

In this section we will describe some results related to the three anomalous 
phenomena discussed in the main text.

\subsection{The Markovian Mpemba effect}\label{SMsect:ME}

\begin{figure}[htb]
\includegraphics[width=8.6cm]{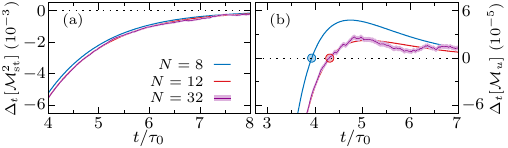}
\caption{Mpemba effect. Evolution of $\Delta_t[\mathcal{A}]$ 
for the observables $\mathcal{A}=\mathcal{M}_\text{st}^2$ (a), and 
$\mathcal{A}=\mathcal{M}_{u}$ (b). We show the results for $N=8$ (blue), 
$N=12$ (red), and $N=32$ (purple, with a lighter shade representing the 
error bars of the MC data). In panel (b), the time at which 
$\Delta_t[\mathcal{M}_{u}]$ changes sign is marked by a dot.}
\label{fig:mpemba_check}
\end{figure}

Figure~\ref{fig:mpemba_check} illustrates the ME as obtained with 
\(T_{b}=1\), \(T_c=4.15\), and \(T_h=15.177\). Let us 
define for simplicity the quantity
$\Delta_t[\mathcal{A}] = ( \mathbb{E}_t^{h}[\mathcal{A}]-
\mathbb{E}_t^{c}[\mathcal{A}])/N$
for any observable. We will consider the two observables not discussed in 
the main text; namely, $\mathcal{A}\in\{\mathcal{M}_\text{st}^2,
\mathcal{M}_u\}$.

In Fig.~\ref{fig:mpemba_check}(a) we see that there is no crossing for 
$\mathcal{M}_\text{st}^2$; this is expected as we have prepared our system 
so that 
$\mathbb{E}_t^{h}[\mathcal{M}_\text{st}^2]$ is exponentially 
accelerated with respect to $\mathbb{E}_t^{c}[\mathcal{M}_\text{st}^2]$.
Hence, the negative sign in Fig.~\ref{fig:mpemba_check}(a).
On the other hand, in Fig.~\ref{fig:mpemba_check}(b), we see a crossing 
similar to those displayed in Fig.~3 of the main text: there is ME 
for the observable $\mathcal{M}_{u}$.

\subsection{Preheating for faster cooling}\label{SMsect:preheating}

\begin{figure}[b]
\includegraphics[width=8.6cm]{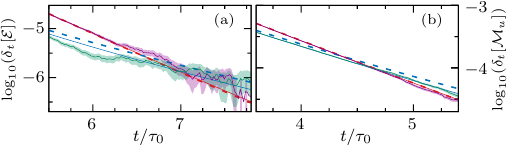}
\caption{Preheating strategy for faster cooling. Evolution of 
\(\delta_t[\mathcal{A}]\) for $\mathcal{A}=\mathcal{E}$ (a),
and $\mathcal{A}=\mathcal{M}_{u}$ (b). Colors red and purple (respectively
blue and green) denote that the protocol includes (respectively does not 
include) an initial 
quench at temperature \(T_{q}\). We show data for $N=8$ (dashed lines), 
$N=12$ (solid lines), and $N=32$ (solid lines with a lighter shade 
representing the error bars of the MC data). Both panels show the expected 
speedup for the preheating protocol, although in (a) the error bars for 
$N=32$ are rather large to draw a definitive conclusion.}
\label{fig:supercooling_check}
\end{figure}

In this section, we compare the one-step protocol against a two-step protocol. 
In the former, we start with system in equilibrium at temperature \(T_0=4.15\), 
and we let this system evolve with a bath temperature \(T_{b}=1\). 
In the latter, we start again at $T_0$, but we first let the system to evolve 
with a bath at temperature \(T_{q}=2000\) up to $t=t_w=0.156$;
at this time, we put the system instantaneously in contact with a another 
bath at temperature \(T_{b}=1\). We expect that this second preheating 
protocol will
make the system to evolve faster than the standard one-step protocol. 
It is now useful to define
$\delta_t[\mathcal{A}] = |\mathbb{E}_t[\mathcal{A}] - 
\mathbb{E}^{T_{b}}[\mathcal{A}]|/N$.

Figure~\ref{fig:supercooling_check} shows that preheating speeds up the 
evolution of the two observables not reported in the main text, namely,  
$\mathcal{A} \in \{ \mathcal{E},\mathcal{M}_{u}\}$. For the energy, 
this behavior is clear for the exact data ($N=8, 12$), but for $N=32$ 
the results after  $t\approx 6.5$ are less definitive due to the large 
error bars of the MC results. 

\subsection{Heating and cooling may be asymmetric processes} 
\label{SMsect:asymmetry}

\begin{figure}[t]
    \begin{tabular}{c}
    \includegraphics[width=8.6cm]{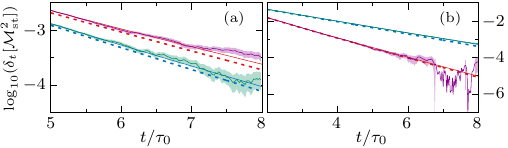} \\
    \includegraphics[width=8.6cm]{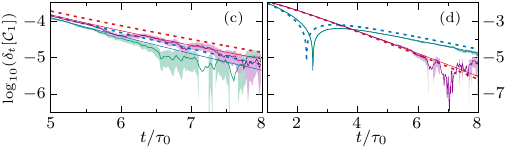} \\
    \includegraphics[width=8.6cm]{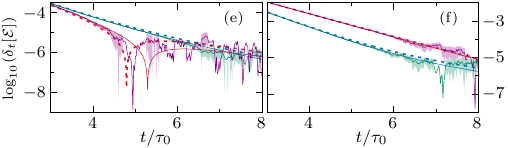} 
    \end{tabular}
    \caption{(A)symmetry in heating and cooling. Evolution of 
\(\delta_t[\mathcal{A}]\) for 
$\mathcal{A}=\mathcal{M}_\text{st}^2$ (a,b), 
$\mathcal{A}=\mathcal{C}_1$ (c,d), and $\mathcal{A}=\mathcal{E}$ (e,f).
In each panel, we depict cooling (blue and green) and heating (red and purple) 
processes. We show data for $N=8$ (dashed lines), $N=12$ (solid lines), and 
$N=32$ (solid lines with a lighter shade representing the errors of the 
MC data). The cooling and heating processes were carried out between: 
(a,c,e) \(T_A=1\) and \(T_B=4.15\); and (b,d,f) \(T_A=15.177\) and 
\(T_B=4.15\). For $\mathcal{M}_\text{st}^2$ and $\mathcal{C}_1$, it is slower 
to approach $T_B=4.15$: i.e., to heat up in (a,c), and to cool down in (b,d). 
However, the behavior is the opposite for the energy (e,f).
}
\label{fig:asymmetry_check}
\end{figure}

Here we will compare the behavior of starting at $T=T_A$ and evolving to the 
bath temperature $T=T_B$ against the inverse process: i.e., starting at 
$T=T_B$ and evolving to the bath temperature $T=T_A$. We have made two 
experiments: one with $T_A=1$ and $T_B=4.15 \approx T_{b}^*$
(see left panels in Fig.~\ref{fig:asymmetry_check}),
and the other one with $T_A=15.177$ and $T_B=4.15$ (see right panels in 
Fig.~\ref{fig:asymmetry_check}). 
Here $T_{b}^*$ is the temperature at which the staggered 
susceptibility $\chi_\text{st}$ attains a maximum 
[see Fig.~2(a) of the main text].

In Fig.~\ref{fig:asymmetry_check} we show the data not reported in the main 
text, and corresponding to the observables 
$\mathcal{A}\in\{\mathcal{M}_\text{st}^2, \mathcal{C}_1, \mathcal{E}\}$ in 
panels (a,b), (c,d), and (e,f), respectively. Note that in panel (d), 
we observe a change of sign at $t\approx 2.5$ in the quantity 
$\mathbb{E}_t[\mathcal{C}_1]-\mathbb{E}^{T_{b}=T_B}[\mathcal{C}_1]$ 
for the system originally at $T_A=15.177$. We also observe such a change 
of sign in panel (e) for the quantity 
$\mathbb{E}_t[\mathcal{E}]-\mathbb{E}^{T_{b}=T_B}[\mathcal{E}]$ at 
$t\approx 5$ for the system originally at $T_A=1$. 

We observe in Fig.~\ref{fig:asymmetry_check}(a,b,c,d), as we did in the main 
text for the observable $\mathcal{M}_{u}$, that for observables  
$\mathcal{M}_\text{st}^2$ and $\mathcal{C}_1$, it is always \emph{faster} 
to move away from $T_B=4.15$ and approach $T_{A}$: i.e., to cool down 
in (a,c), and to heat up in (b,d). 

However the behavior for the energy $\mathcal{E}$ is the opposite [see 
Fig.~\ref{fig:asymmetry_check}(e,f)]: it is faster to approach $T_B=4.15$: 
i.e., to heat up in (e) and to cool down in (f). The explanation of this 
effect was already given in the main text: for bath temperatures close 
to $T_{B}=4.15 \approx T_{b}^*$, the coefficient 
$\beta_2^\mathcal{E} \approx 0$ [cf. Eq.~\eqref{SMeq:DD3a}].

%%%%%%%%%%%%%%%%%%%%%%%%%%%%%%%%%%%%%%%%%%%%%
\bibliographystyle{apsrev4-1}
%\bibliography{Ising1D.bib}